# Capabilities and limitations of pure-shear based macroscopic forming simulations for 0°/90° biaxial non-crimp fabrics

Bastian Schäfer[1,a*], Dominik Dörr[2,b], Naim Naouar[3], Jan Paul Wank[1], and Luise Kärger[1,c*]

[1]Karlsruhe Institute of Technology (KIT), FAST - Lightweight Engineering, 76131 Karlsruhe, Germany

[2]Simutence GmbH, 76131 Karlsruhe, Germany

[3]CNRS, INSA Lyon, LaMCoS, UMR5259, 69621 Villeurbanne, France

[a]bastian.schaefer@kit.edu, [b]dominik.doerr@simutence.de, [c]luise.kaerger@kit.edu

**Abstract.** Macroscopic modeling of a non-crimp fabric's (NCF's) forming behavior is challenging as it strongly depends on the textile architecture, fiber type, and stitching type. While shear is the main deformation mode of woven fabrics, membrane modeling approaches for NCFs should also consider stitching deformation and roving slippage. However, for 0°/90° biaxial NCFs (Biax-NCF) with a symmetrical stitching pattern and high stitch pretension, deviations from a pure-shear assumption in coupon tests are only observed at higher shear angles due to limited roving slippage. In this work, a hyperelastic approach initially proposed for unidirectional NCFs is adopted for a tricot stitched 0°/90° Biax-NCF based on a pure-shear assumption. The shear behavior is experimentally characterized through 45° off-axis-tension tests, and the parameterization is derived from energetic approaches originally developed for woven fabrics. This approach efficiently and adequately describes the general behavior in forming simulations of different geometries. Fiber orientation and location of areas with high shear angles are predicted well, but the peak shear angles are overestimated due to the neglected roving slippage.

**Introduction**

Macroscopic forming simulations of engineering textiles can be used to efficiently investigate and subsequently optimize manufacturing processes of continuous fiber reinforced plastics. To accurately predict the forming behavior and potential forming defects, the modeling of a textile must consider its main deformation modes, which are strongly dependent on its mesoscopic architecture. Non-crimp fabrics (NCFs) consist of one (UD-NCF), two (Biax-NCF) or more stacked layers of unidirectional plies of continuous fiber rovings that are interconnected by an intricate stitching pattern to effectively avoid weakening undulations of the fiber rovings. Thus, the architecture, fiber type, and stitching pattern of an NCF [1] must be accounted for when selecting the modeling approach to describe the membrane behavior.

As with other continuous reinforcing fabrics, the tensile strains of the rovings are limited during forming, and the shape change for double-curved geometries predominantly occurs through other deformation modes. In UD-NCF [2,3], these are large shear, superimposed with perpendicular in-plane compression of the rovings, and transverse tensile strains due to stitching deformation as well as relative roving slippage. In Biax-NCF with perpendicularly oriented fibers, the transverse strains of both fiber layers are mutually constrained by the second fiber orientation and depend on the stitching pattern as well as pretension [4]. For ±45° Biax-NCFs with the principal direction of the stitching pattern offset from the fiber directions, the tensile stiffness of the stitching must be considered as an independent deformation mode due to a resulting asymmetric material behavior under shear [5-8]. In contrast, 0°/90° Biax-NCFs display varied behavior depending on stitch pretension. Low stitch pretension allows significant slippage between the fiber layers even with small deformations, necessitating two element layers combined with a suitable inter-ply model in macroscopic approaches [9]. Conversely, high stitch pretension results in symmetrical material behavior, with deviations from pure-shear assumptions occurring only at higher shear angles due to limited roving slippage [10-12].

In this study, the capabilities and limitations of a pure-shear assumption for a symmetrical 0°/90° biaxial NCF with high stitch pretension are investigated. A hyperelastic approach, originally proposed for UD-NCF [3], is adopted for a tricot stitched 0°/90° biaxial NCF based on a pure-shear assumption. The idea of applying this approach for Biax-NCF was first proposed in the thesis by Schäfer [13]. The shear behavior is characterized experimentally using 45° off-axis-tension tests [12], and the parameterization is derived from energetic approaches originally developed for woven fabrics [14]. The bending behavior in warp and weft direction is characterized with cantilever tests. The performance of the approach is evaluated by comparison with experimental strain measurements from 30° off-axis-tension tests as well as forming tests for different initial orientations and geometries (hemisphere, tetrahedron, square box).

**Pure-shear based macroscopic forming simulation of Biax-NCF**
Investigated material. In this study, a nearly balanced Biax-NCF without binder called MD600 manufactured from carbon fibers by Zoltek™ is investigated, cf. Fig. 1 a). The fabric with an initial thickness of $t_0 = 1.004$ mm consists of two layers of fibers, each weighing approximately 300 g/m². They are sewn together in a 0°/90° orientation in a tricot stitching pattern.

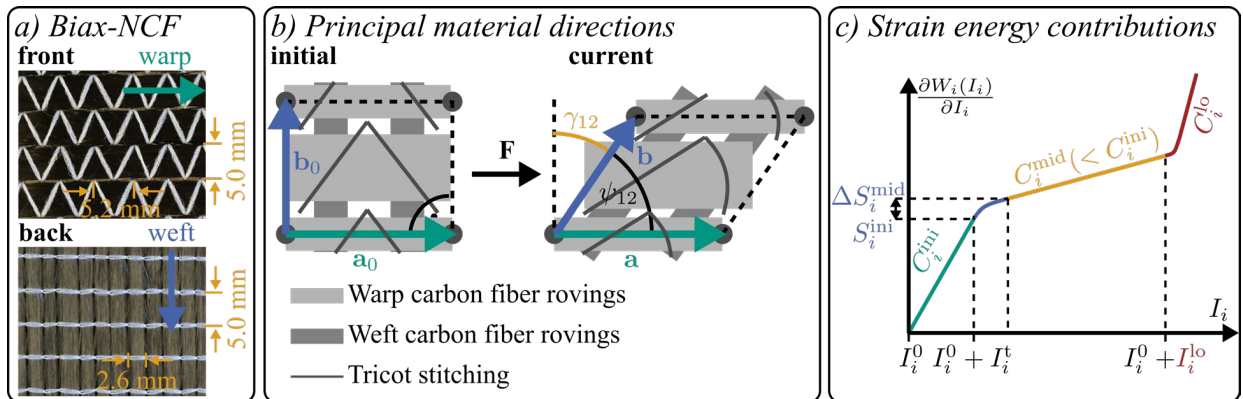

Figure 1 – a) Biaxial 0°/90°-NCF (Biax-NCF); b) Principal material directions in the initial and current configuration; c) Generalized strain energy density contribution for a typical three-stage stress curve [3].

Hyperelastic membrane model. The hyperelastic membrane approach applied in this study was initially proposed for macroscopic modeling of UD-NCF [3] and is adapted in this study for Biax-NCF based on a pure-shear assumption. The approach is implemented by means of a user-defined material behavior (VUMAT) and applied to triangular membrane elements (M3D3) in ABAQUS/EXPLICIT. The element edges are aligned with the initial fiber orientations in all models to alleviate numerical intra-ply shear locking [14].

Tension in both roving directions and shear deformation are considered, while relative roving slippage is neglected. Therefore, pseudo-invariants $I_i$ based on the principal material directions $\boldsymbol{a}_0$ and $\boldsymbol{b}_0$ are introduced, cf. Fig. 1 b, Eqs. 1 to 5),

$$I_4 = \boldsymbol{a}_0 \cdot \boldsymbol{C} \cdot \boldsymbol{a}_0, \tag{1}$$

$$I_6 = \boldsymbol{a}_0 \cdot \boldsymbol{C} \cdot \boldsymbol{b}_0, \tag{2}$$

and

$$I_8 = \boldsymbol{b}_0 \cdot \boldsymbol{C} \cdot \boldsymbol{b}_0, \tag{3}$$

with

$$I_4^0 = I_8^0 = 1 \tag{4}$$

and

$$I_6^0 = 0, \tag{5}$$

where $\boldsymbol{C}$ is the right Cauchy-Green Tensor, $I_4$ and $I_8$ correspond to the quadratic stretches in warp and weft direction, respectively, and $I_6$ is the shear strain related to the shear angle $\gamma_{12}$ by Eq. 6:

$$\gamma_{12} = \arccos(a_0 \cdot b_0) - \arccos\left(\frac{I_6}{I_4 \cdot I_8}\right). \tag{6}$$

The total strain energy density function for the resulting stress of the Biax-NCF [13] is additively decomposed into (Eq. 7):

$$\boldsymbol{S} = 2\frac{\partial W_{\text{tot}}^{\text{Biax}}(\boldsymbol{C})}{\partial \boldsymbol{C}} = 2\sum_i^N \frac{\partial W_{\text{tot}}^{\text{Biax}}}{\partial I_i}\frac{\partial I_i}{\partial \boldsymbol{C}}, \tag{7}$$

with:

$$W_{\text{tot}}^{\text{Biax}}(I_4, I_6, I_8) = W_4(I_4) + W_6(I_6) + W_8(I_8). \tag{8}$$

A generalized piecewise function for the derivative of the strain energy density [3], cf. Fig. 1 c), is applied for the individual contributions by Eq. 9:

$$\frac{\partial W_i(I_i)}{\partial I_i} = \begin{cases} C_i^{\text{ini}} I_i^a, & [0 \leq I_i^a < I_i^t] \\ S_i^{\text{ini}} + \left(C_i^{\text{mid}}(I_i^a - I_i^t) + \Delta S_i^{\text{mid}}\right) \cdot \left(1 - e^{-c_i^t(I_i^a - I_i^t)}\right), & [I_i^t \leq I_i^a < I_i^{\text{lo}}] \\ S_i^{\text{ini}} + \left(C_i^{\text{mid}}(I_i^a - I_i^t) + \Delta S_i^{\text{mid}}\right) \cdot \left(1 - e^{-c_i^t(I_i^a - I_i^t)}\right) + C_i^{\text{lo}}(I_i^a - I_i^{\text{lo}})^2, & [I_i^{\text{lo}} \leq I_i^a] \end{cases} \tag{9}$$

with $I_i^a = |I_i - I_i^0|$, $I_i^t = S^{\text{ini}}/C^{\text{ini}}$ and $c_i^t = C^{\text{ini}}/\Delta S_i^{\text{mid}}$. The generalized function can represent the typical three-stage response during shear or elongation, as observed in previous studies [2,4,7], with six physically interpretable parameters related to the individual stages, cf. Fig. 1 c). The initial stiffnesses in warp and weft direction are chosen as sufficiently high constant values ($C_4^{\text{ini}} = C_8^{\text{ini}} = 1000$ MPa), instead of a numerically complex inextensibility condition. This assumption efficiently limits the roving deformation during forming, by preventing significant tensile strains. However, this also prevents the consideration of relative roving slippage. To model roving slippage in macroscopic approaches with a single element layer, positive tensile strains would be required [15].

Shear parameterization. Experimental off-axis-tension tests (OATs) with an initial length of $l_0 = 320$ mm, width of $w_0 = 160$ mm and offset angle of $\Theta_0 = 45°$ are used to parameterize the shear behavior, cf. Fig. 2 a). The experimental results are detailed in Schäfer et al. [12]. The second Piola-Kirchhoff shear stress $S_{12}$ is incrementally calculated based on the energetic approach for the theoretical deformation proposed for woven fabrics by Boisse et al. [14] according to Eq. 10:

$$S_{12}(\gamma) = \frac{f(\gamma)(l_0 - w_0)\left(\cos\left(\frac{\gamma}{2}\right) - \sin\left(\frac{\gamma}{2}\right)\right)}{w_0^2 t_0 \cos(\gamma)} - \frac{\cos\left(\frac{\gamma}{2}\right)}{2\cos(\gamma)} S_{12}\left(\frac{\gamma}{2}\right), \tag{10}$$

where $f$ is the machine force and $\gamma$ is the shear angle measured in the specimen's center via digital image correlation (DIC). The resulting experimental shear stress is shown in Fig. 2 b). For large shear angles ($\gamma > 50°$), the forces increase only moderately compared to woven fabrics, as strong fiber slippage is observed in the transition zones between the full-shear ($\gamma$) and half-shear ($\gamma/2$) region [12]. Two different parameter sets are discussed in the following Sections, cf. Table 1.

*Table 1 – 'Best-fit' and 'locking' material parameter sets of the hyperelastic membrane approach for Biax-NCF.*

|  | $C_i^{\text{ini}}$ [MPa] | $S_i^{\text{ini}}$ [MPa] | $C_i^{\text{mid}}$ [MPa] | $\Delta S_i^{\text{mid}}$ [MPa] | $C_i^{\text{lo}}$ [MPa] | $I_i^{\text{lo}}$ [−] |
|---|---|---|---|---|---|---|
| $W_4$ | 1000 | − | − | − | − | − |
| $W_8$ | 1000 | − | − | − | − | − |
| $W_6^{\text{best-fit}}$ | 0.334 | 0.00386 | 0.00196 | 0.0399 | 0.911 | 0.610 |
| $W_6^{\text{locking}}$ | 0.334 | 0.00386 | 0.00196 | 0.0399 | 1000 | 0.819 |

First, the generalized approach, cf. Eq. 9, is fitted to the experimental shear stress with the Nelder-Mead algorithm to determine a 'best-fit' parameter set, cf. Fig. 2b). An adequate agreement is achieved until a shear angle of $\gamma \approx 57°$. For higher shear angles, the locking behavior is underestimated due to the limited complexity of the generalized approach function originally proposed in Schäfer et al. [3] to describe UD-NCF with as few material parameters as possible.

Second, a 'locking' parameter set is proposed, to induce a significant locking for $\gamma > 57°$ by manually modifying the locking onset $I_6^{\text{lo}}$ and stiffness $C_6^{\text{lo}}$, cf. Eq. 9. The locking onset was chosen based on the maximum measured shear angle during the forming tests [16], and to prevent a collapse of highly sheared elements during the simulation of the off-axis-tension tests. The remaining parameters remain the same to compare both parameterizations. Modifications to the stress function itself were avoided to investigate the capabilities of the original hyperelastic model with its intentionally limited complexity. In the future, however, modifications to the generalized approach function could be explored to combine both parameterizations.

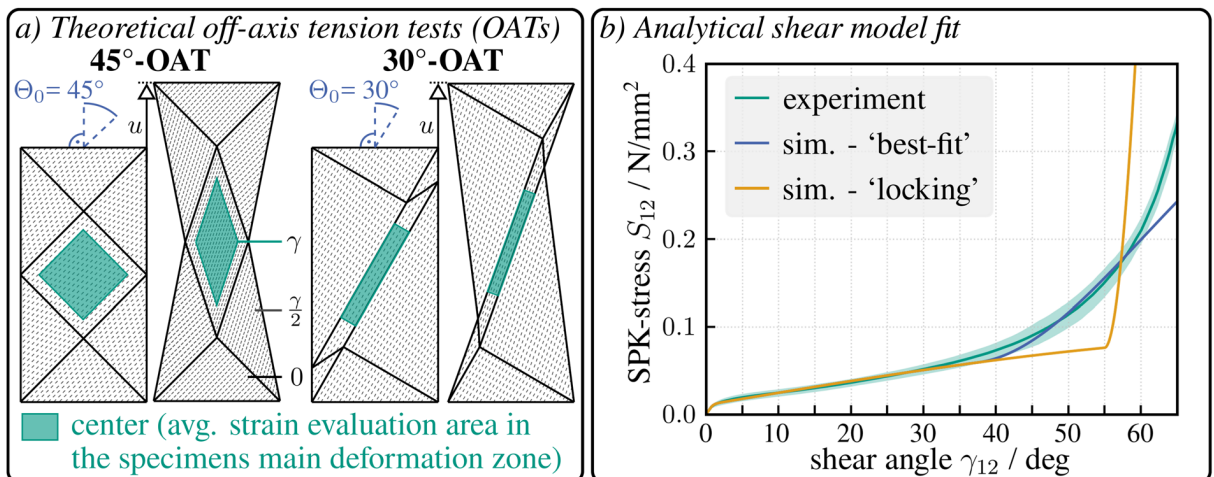

*Figure 2 – a) Theoretical deformation of the 45°- and 30°-off-axis-tension tests (OATs) [adapted from 12]; b) Fitting of the shear energy density contribution in the simulation to the experimental results to determine a 'best-fit' and a 'locking' parameter set for the simulation.*

**Numerical study 1: Off-axis tension tests**

The membrane material model with the two different parameter sets outlined above is applied to simulations of 45°- and 30°-OATs, cf. Fig. 2 a), to verify the parameterization strategy and to investigate the limitations of assuming pure-shear and neglecting relative fiber slippage. The resulting strains in the specimen's center and forces are shown for both parameter sets and compared to experimental results in Fig. 3 a).

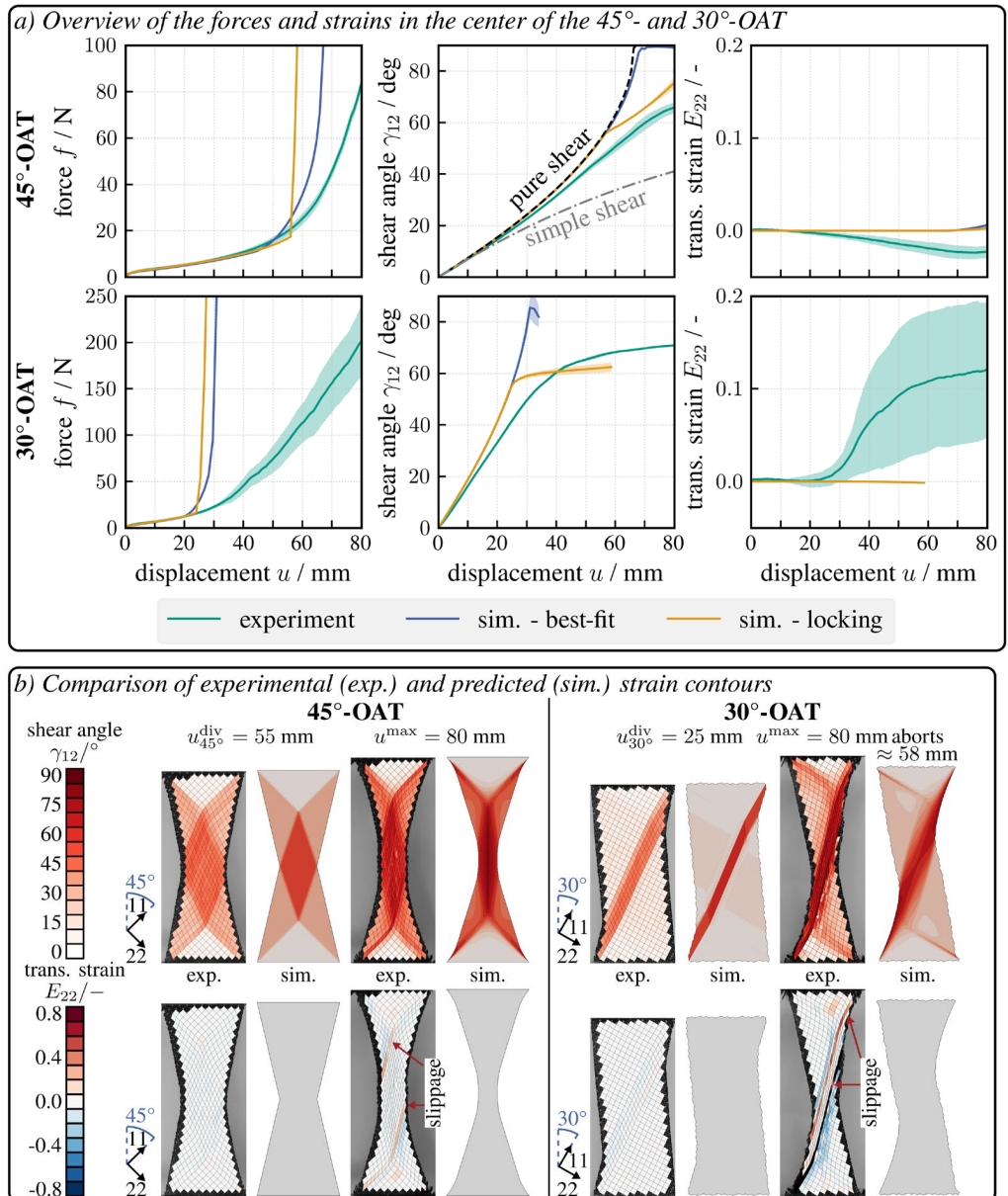

Figure 3 – a) Overview of forces and averaged strains in the centre of the 45°- and 30°-OATs; b) Resulting strain contours for the shear angle and transverse strains with the 'locking' parameter set [adapted from 12, 13].

For both parameter sets, the predicted shear angle in the 45°-OAT initially aligns with the pure-shear assumption for displacements $u < 30$ mm. However, relative roving slippage is observed in the experiments, which is indicated by positive transverse tensile strains along the edges of the main deformation zone, cf. Fig.3 b), as well as $E_{22} \neq 0$ for the local strains in the center. Thus, the experimentally measured shear angle in the center is smaller, leading to a significant overestimation of the predicted forces and shear strains at higher displacements. A similar trend is evident in the 30°-OAT, where force overestimation occurs at even lower displacements.

In simulations with the 'best-fit' parameter set, elements in the center collapse completely at a shear angle of 90°, causing the simulation of the 30°-OAT to abort at $u \approx 30$ mm. At this displacement, significant transverse tensile strains are measured during the experiments in the specimen's center, cf. Fig. 3 a), as well as along the edges of the main deformation zone, cf. Fig.3 b). In contrast, the 'locking' parameter set exhibits an earlier and more pronounced onset of locking in the 45°-OAT, which restricts the achievable shear angle and postpones the complete collapse of elements in the 30°-OAT until $u \approx 58$ mm.

Consequently, the neglected roving slippage in the simulation approach makes it impossible to model both the 30° and 45° OAT for large displacements, regardless of the chosen parameters. For subsequent analyses, the 'locking' parameter set is adopted as it shows better agreement with experimental shear angles and improved numerical stability at large displacements.

Fig. 3 b) shows the strain contours of the simulation for the 'locking' parameter set in the 30°- and 45°-OAT at the displacements $u_{30°}^{\text{div}} = 25$ mm and s $u_{45°}^{\text{div}} = 55$ mm, where the predicted forces diverge significantly from experimental results, and the maximum displacement $u^{\max} = 80$ mm. At displacements below $u^{\text{div}}$, the orientation and position of the shear zones are in good agreement between simulations and experiments. However, the inability of the simulation approach to account for roving slippage ($E_{22} \neq 0$ in the experiments) limits the accuracy of the predictions at higher displacements.

**Numerical study 2: Forming simulations of different geometries**

Forming test setup. The proposed approach is applied to forming simulations with three different punch shapes with initial layer orientations of $\Theta_0 = 0°$ and $\Theta_0 = 45°$. A hemisphere, a tetrahedron and a box geometry, cf. Fig. 4, are investigated to further evaluate the performance of the model for different geometrically challenging features and to identify limitations of a pure-shear assumption for Biax-NCF. For all geometries, planar blank holders apply their own weight $m$ to enable 2D strain measurements via DIC in the flat areas. Global strain contours were measured for all tests. Additionally, locally averaged shear angles $\bar{\gamma}$ were calculated in three distinct areas (green, blue and orange) for each geometry, cf. Fig. 4. The experimental procedure and results are detailed in Schäfer et al. [16].

Bending modeling. The bending behavior of Biax-NCF must be modeled decoupled from its membrane behavior due to the fabric's fibrous nature with low transverse shear stiffness. Thus, an existing hypoelastic bending model originally developed by Dörr et al. [17] for thermoplastic tapes is used, which was also successfully applied for woven fabrics [18]. The approach enables accurate modeling of the bending stiffness in warp and weft direction also under large shear deformation. The bending stiffnesses were determined in standard Peirce cantilever tests to $B_{11}^{\text{warp}} = 3.403$ Nmm and $B_{22}^{\text{weft}} = 3.863$ Nmm. The approach is implemented as a user-defined through-thickness integration scheme (VUGENS) for S3R shell elements, which are superposed to the membrane elements via shared nodes.

Friction modeling. The general contact algorithm available in ABAQUS/EXPLICIT is applied to model friction between the planar blank holders and the fabric. The average friction coefficient of $\mu = 0.157$ was determined in sled-based pull-over tests, which are detailed in Schäfer et al. [19].

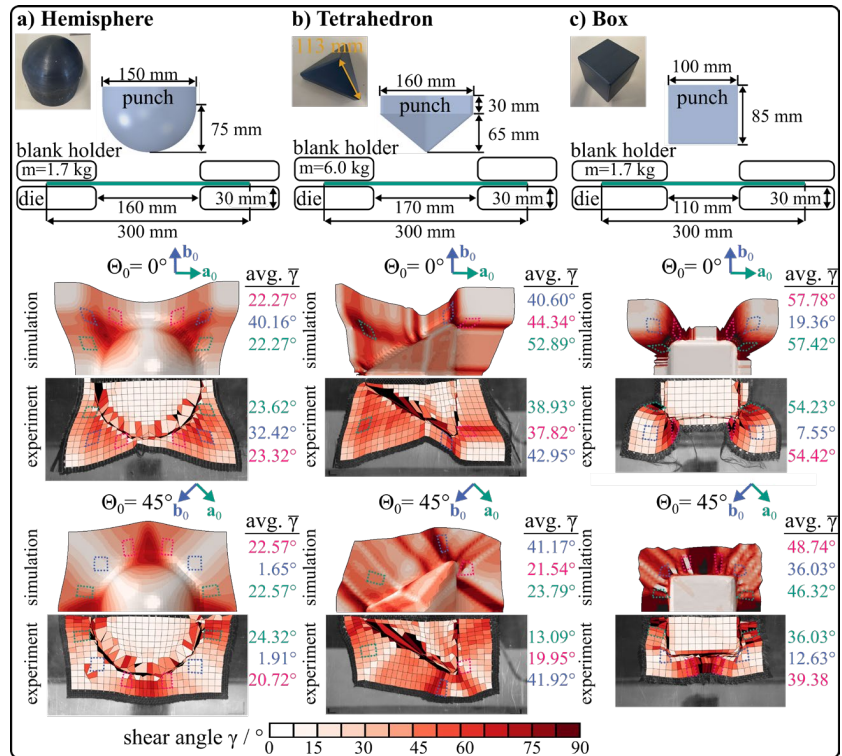

*Figure 4 – Schematics of the test setups [adapted from 16] as well as comparison of experimental and predicted shear angle contours for [0°] and [45°] single-layer configurations for the a) hemisphere, b) tetrahedron and c) square box forming tests.*

Simulation results. The experimental and predicted shear angle distributions after forming for the three punch shapes are shown in Fig. 4. The blank's outer contours can be used to evaluate the overall deformation behavior based on the material draw-in [3,6]. The simulations are in good qualitative agreement with the experiments. The general location and shape of areas with high shear angles are predicted well for all configurations. This indicates an adequate modeling of the Biax-NCF's general deformation behavior by the pure-shear approach. However, for some configurations, the simulation overestimates the peak shear angles in highly deformed areas. These observations are similar to the conclusions for the off-axis tension tests, cf. Fig. 3.

For a more detailed comparison, the average shear angles in three distinct local areas (green, blue and pink) are compared. In many areas an acceptable quantitative agreement ($\Delta \bar{\gamma} < 10\%$) with the experiments is achieved. However, for some configurations, the local shear angle is overestimated in zones further away from the punch area. The deviations ($\Delta \bar{\gamma} > 10\%$) are especially notable for the blue area in the 0°-hemisphere test, the pink area for the 0°-tetrahedron test, the green area for both tetrahedron tests, the blue area for 0°-box test, and all areas in the 45°-box-test. This is attributable to relative roving slippage within the fiber layer and between both fiber layers of the Biax-NCF, which is identified as positive transverse tensile strains in these areas during the experiments [16].

In the forming simulations for the box punch, local shear bands occur in the flat areas between the diagonal flaps for $\Theta_0 = 0°$, as well as in the diagonal areas for $\Theta_0 = 45°$. The shear bands result from numerical instabilities, due to the very high shear angles above 60° combined with small out-of-plane wrinkling in the gap between the die and the punch. Similar behavior is also observed by Bai et al. [20] in box forming experiments on woven fabrics with comparable material properties and simulations with a pure-shear based approach.

**Summary and discussion**
In this study, a macroscopic forming model previously developed for UD-NCF was adapted for a balanced 0°/90° Biax-NCF based on a pure-shear assumption, thereby demonstrating the general transferability of the hyperelastic approach. The model was efficiently parametrized using an energetic method developed for woven fabrics, providing a good approximation of the investigated Biax-NCF's behavior for moderate deformations. However, the macroscopic model's inability to account for roving slippage limits its accuracy at higher strains. Instead, an artificial shear locking was introduced to increase stability and prevent the collapse of elements for high shear angles.

The proposed approach is capable of an adequate prediction of the global forming behavior for different initial fiber orientations and punch shapes with different geometric complexity. The outer contour after forming as well as the location and shape of highly sheared areas are in good agreement with experimental results. Nevertheless, peak shear angles are overestimated, reflecting the model's limitations due to the exclusion of roving slippage.

In conclusion, the presented method provides a decent approximation of the forming behavior of Biax-NCF based on pure-shear, with a fast and efficient parameterization process. It can therefore be used for prediction of critical areas during forming to assist in process development and optimization. However, to improve prediction accuracy for large deformations, future developments should incorporate methods to account for relative roving slippage. For macroscopic approaches with a single element-layer, methods that locally reduce the tensile stiffness in regions of high shear angle gradients, as proposed by Wank et al. [15], could be used to describe roving slippage via tensile strains. Alternatively, macroscopic approaches with two-element layers, as suggested by Bel et al. [9], could be used to describe the relative slippage within a fiber layer with an approach for UD-NCF and enable interlayer slippage using a suitable contact formulation.

**Acknowledgement**
The authors would like to thank the German Research Foundation (DFG) and the French National Research Agency (ANR) for funding the collaborative project 'AMECOMP' (DFG: 431354059, ANR: ANR-19-CE06-0031), for which the work was initiated, and the German Federal Ministry for Economic Affairs and Climate Action (BMWK) for funding the LuFo project 'Electra' (20W1912D), for which the work was continued. This work is also part of the Heisenberg project 'Digitalization of fiber-reinforced polymer processes for resource-efficient manufacturing of lightweight components', funded by the DFG (455807141).